\documentstyle[12pt]{article}
\begin{document}
\title{Complete Entanglement-Based  Communication with Security.} 
\author{Arindam Mitra
\\Lakurdhi, Tikarhat Road,Burdwan, 713102\\
West Bengal,  India.}

\maketitle

\begin{abstract}\bf 
\noindent If, due to some catastrophe, our classical communication system is 
destroyed but quantum entangled  state survived, in that entanglement age,
it is widely believed,  the whole world  would be communication less.
It is discussed that complete entanglement based communication even with
security is possible in that scenario. In this context, the notion of  cheating-free 
Bell's inequality test is introduced.

\end{abstract}

\newpage
\noindent
Not long before, entanglement, the enigmatic feature of quantum correlation [1,2],
 was believed to be useful only to  address the interpretational aspect of quantum mechanics.
But things have changed after the discovery of quantum cryptography [3-5],
quantum computation [6-7], dense coding [8] and teleportation [9].
Now entanglement is the part and parcel of quantum information.\\

Deutsch, in his foundational  paper on quantum computation, 
obliquely suggested [6] 
that  quantum entanglement can be used to generate quantum key, discovered
by Bennett and Brassard [4].
Bennett and Brassard used disentangled state in
their BB84 quantum key distribution (QKD) scheme, so Deutsch's suggestion was striking.
Ekert elaborated Deutsch's suggestion and pointed out [5]
Bell's theorem [2] can be used to test eavesdropping. Although
practical utilization
of Bell's theorem is conceptually interesting but it has been soon clear [10]
that avoiding Bell's inequality test one can use EPR correlation [1]
to detect eavesdropping. Be that as it may, finally it has been settled that
EPR and non EPR quantum cryptosystems (i.e. entanglement and disentanglement
based systems) are two different kind of system however they have same
physical property - the ability to expose clandestine eavesdropping.\\

Bennett, Brassard and Mermin  carried out a comparative study [10] on
entanglement and disentanglement based  quantum cryptosystems.
They concluded that only  difference is that  random
data are jointly chosen by legitimate users in disentanglement based  system but
in entanglement based system,
random data  spontaneously originates due to the  measurements.
In entanglement based cryptosystem key does not exist - not even in the mind of legitimate users
until measurement is performed. If sender has no control
over the random data arising from  entanglement based cryptosystem then 
it can be assumed that  
completely entanglement based secure communication is impossible. \\

But if we are not interested about security
can we get completely entanglement based communication ?
Such possibility is recently ruled out by  Bennett and Divincenzo [11].
They assume:
"entanglement by itself can not be used to
transmit a classical message". It means
entanglement could not produce meaningful data, which is needed to send
a classical message over an entirely entanglement based channel. 
We shall see the above  assumptions
do not  hold good for  our  information processing   technique [12]
because entanglement can be manipulated to produce meaningful data. That is, 
the power of entanglement
in quantum information processing is surprisingly underestimated.\\

Let us recall the technique. The two different sequences of quantum
states (say, $S_{0}$ and $S_{1}$) represent bit 0 and  1. The information regarding 
 two sequences  is initially secretly shared between the
legitimate users. Repeatedly and randomly using these two sequences 
arbitrarily long string of
bits can be generated. Now we shall use entangled state to
prepare the two sequences. \\

Suppose Alice has a personal database which contains meaningful data.
She wants to transmit the database to her partner Bob stationed at
far off distance only using entangled states. Suppose  Alice possesses a
stockpile  of EPR pairs. The n pairs (n is a moderately large number) can be arranged in two
different   ways  to represent bit 0 and 1.\\ 

\noindent
Suppose $ S_{1} = \left\{A,\, B,\, b,\, C,\,  D,\, a,\, E,\, F,\, e,\, f,\, d,\,
 c,......\right\} $  \\
and $ S_{0} = \left\{A,\,  B,\, C,\, c,\, D,\, a,\, b,\, E,\, F,\, e,\, f,\, 
d,......\right\} $.\\
Here "A-a", "B-b", "C-c", "D-d", "E-e" and"F-f" 
denote  EPR pairs.

\paragraph*{} The quantum key $K_{Q}$ will be :  
\begin{eqnarray}\left(\begin{array}{ccccccccccccc}
A & B & b & C & D & a & E & F & e &  f & d & c &  ....\\
A & B & C & c & D & a & b & E & F &  e & f & d &  ....\\
A & B & C & c & D & a & b & E & F &  e & f & d &  ....\\
A & B & b & C & D & a & E & F & e &  f & d & c &  ....\\
. & . & . & . & . & . & . & . & . &  . & . & . &  ....\\ 
. & . & . & . & . & . & . & . & . &  . & . & . &  ....\\ 
. & . & . & . & . & . & . & . & . &  . & . & . &  ....\\ 
A & B & C & c & D & a & b & E & F &  e & f & d &  ....

\end{array}\right)
\equiv \left(\begin{array}{c}
1 \\ 0 \\ 0 \\ 1 \\. \\. \\.\\ 0\end{array}
 \right)\nonumber\end{eqnarray}

The quantum state of the pairs can be describes as,\\
$\vert \psi\rangle_{i,j=1}^ {2n} =
1/{\sqrt 2} (\vert{\uparrow}\rangle_{i}^{1}\vert{\downarrow}\rangle_{j}^{2}
-\vert{\downarrow}\rangle_{i}^{1}\vert{\uparrow}\rangle_{j}^{2})$, 
where i and j ($i \neq j$) denote the position of any
pair in the two arrangements which are not very close to each other.
The information regarding the two  arrangements ($S_{0}$ and $S_{1}$)
is initially secretly shared between them. \\

If Alice wants to send bit 0 from her database, she arranges n pairs
according to $S_{0}$ and  send that sequence to Bob. Similarly she can send bit 1 by
sending $S_{1}$. 
Bob can easily recover the bit values from the incoming sequences by correlating
the shared information regarding the two probable arrangements with the outcome of his
sequence of measurements.\\

Let us describe  a simple recovery of the bit value
assuming (for clarity) only one type of the sequence is sent by Alice.
Bob first receives the sequence of EPR  particles and stores them in a quantum
memory-cum-register. He sorts out $n/2$ pairs assuming they belong to $S_{0}$
and keep the  $n/2$  pairs in $(n/2) \times 2$ memory array marked 0.
The remaining n/2 pairs (say n is even number)are assumed to belong to $S_{1}$ and kept
in $(n/2) \times 2$ memory array marked 1.
At this stage she does not know which identification is right.
He measures the spin in vertical direction on EPR pairs (
each pair is kept in a row). If the sequence is $S_{0}$, then the results
corresponding to  rows of the array marked 0 will be either $\uparrow$ and 
$\downarrow$ 
or
$\downarrow$ and $\uparrow$.  But the results, corresponding
to the rows of array marked 1, will be four types; 1) $\uparrow$ and $\downarrow$, 2) $\downarrow$ and
$\uparrow$, 3)
$\uparrow$ and $\uparrow$ 4) $\downarrow$ and $\downarrow$.
As array marked 0 only contains EPR data so bit 0 is recovered. If $S_{1}$
is sent then array marked 1 will contain EPR data.  Thus Alice
can sent bit 0 or 1 according to her wish and Bob can recover the bit values. 
It means the 
complete
entanglement based communication is possible. Next we shall discuss how
this communication can be made secure. \\

Eavesdropper's problem is identical to what she encountered in our
disentanglement based QKD scheme [12]. He/she can not extract the bit value 
from a single copy of
any sequences [13].  As our scheme is based on repetition, eavesdropper,
extracting bit
values, can evade detection if proper security criterion is not 
imposed. The security criterion is:  bit by bit security. In this criterion,
Alice  will  send the next bit after being informed by Bob that
the previous one has not been 
corrupted by eavesdropper. This test needs two-way communications 
which simultaneously give  authentication [12].
If the bit is not corrupted Bob can inform Alice by sending
any of the two sequences.
It is not necessary to send back the same bit value sequence what
Bob has got. \\  

Note that, if security is not necessary, initial sharing of information
is also not needed. Bob can
recover the message by correlating the results of randomly incoming
identical sequences representing identical bit values.  The concept of secret sharing of random data
 was first introduced by
Vernam in classical cryptography [14]. But the problem of that classical
code is that the same shared information can not be again and again used. 
On the other hand, in our scheme, repetition of 
the  quantum sequence is simply possible
because each  sequence can be made secure by 
the no-cloning/uncertainty
principle.\\
 
The above protocol can  also be used as 
three party protocol involving Alice (sender),
 Bob and Sonu (receivers). In three party protocol
each of the two receivers - Bob and Sonu -  will 
get one of the EPR particles of each  pair belonging to 
any  of the two arrangements. 
Let us take an example.\\ 

\noindent $ S_{0}^{Bob}
= \left\{ A, \,\,B,\,\, C,\,\, D,\,\, E,\,\, F,\,\, G,\,\,,......\right\} $\\
\noindent $ S_{0}^{Sonu} = \left\{ b,\,\, g,\,\,a\,\, c, \,\,d,\,\, f,\,\, e
.......\right\} $.\\

The above two arrangements are
representing bit 0. Bob is  given the first arrangement
and  Sonu is given the second arrangement.
If they co operate, they could 
recover the bit 0.  
Similarly their co-operation 
will be required to recover the bit 1.
This is actually entanglement based alternative 
message splitting [15,16] protocol.\\

The encoding is  described for two-particle
entangled state, although many-particle entangled state
can be used. So far security is  concerned which type of
entangled state should be chosen in  a sequence of n particles ?
For example: 200  three-particle or 300 two-particle entangled
states ? Next we shall discuss that the question   is related  to an
ignored part  of security of quantum encryption. \\

It is said that in quantum encryption, eavesdropper is bound to
introduce errors  due to no-cloning principle. In strict sense,
this statement is incorrect. Eavesdropper has nonzero probability 
( but extremely small) of success in guessing the entire encoding. In other words, we can
say that  quantum states, used for encoding, can be perfectly cloned with
non-zero probability by the eavesdropper. Hence eavesdropper can evade
detection with some non-zero probability. In quantum fashion, the success
of perfect guess can be described  as an entanglement of eavesdropper's
mind with  users' minds.  After all quantum mechanics cannot resist
two mind beating in unison for  a while, since law of probability
allows it. If we think the existence of many many eavesdroppers then 
this  guess-strategy may work for one of the eavesdroppers. Then there
will be no errors due to the measurements of that eavesdropper. So,
we should protect the system against such wild strategy. This is possible
because probability of success depends on how many different ways a
particular encryption can be executed. For our encryption it simply depends
on the permutation of the quantum states. Therefore it is easy to enhance
the  security (say, inner layer of security) of our encryption.
Let us see how many different ways the particles can be arranged.
If n is the total number of particles in each  sequence and r is the
number of particles in an entangled state,  the number of distinguishable
arrangements (they can be distinguished by measurements if many copies of 
the same arrangement are given) is $n!/r!^{n/r}$,
where n is a factor of r. Eavesdropper's chance of correct guess is
$p_{r}=r!^{n/r}/n!$. Now we shall see that for fixed n, two-particle
entangled state is the best choice to enhance the inner layer of security.\\

Suppose n particle sequence is composed either by x copies of two-particle 
entangled states or y copies of r-particle entangled states.
Assume,  two-particle is the best choice. Then  
$p_{r}/p_{2} = r!^{y}/2^{x} =r!^{2x/r}/2^{x} > 1$.
It follows 2 log r!> r log 2.  This identity is always true. 
So two-particle is the best choice. Note that, in classical encryption
we do not have this kind of choice. Choice is made due to the 
indistinguishability of quantum particles forming entanglement. 
For  our encoding the issues like channel capacity, entropy and 
statistical distinguishability can be further investigated. \\

From Ekert's work it is known that EPR based cryptosystem can be protected
by Bell's theorem. But the problem  is: 
violation or no-violation of Bell's 
inequality is not necessarily mean a genuine test of entanglement.
Using disentangled states (BB-84 states[2]) 
Bell's inequality can  be tested[17].
By standard meaning, this is a  fake Bell's inequality test.
Suppose Bell's inequality is tested by two distant experimentalists.
In that case, it is widely believed [18,20] that there is no way to know 
whether the test is real or fake because there is no way to
know whether they share genuine entangled state or not.
If it be so,
 experimental falsification
of  hidden variables  under Einstein's locality
condition [1] is perhaps incomplete to an experimentalist
because he/she can be cheated by  other 
remote experimentalist participating
in that test [19]. In the cheating case,
pseudo hidden variable takes the place of
hypothetical local hidden variable.
Next we shall discuss that cheating-free
test is simply possible.  \\

The problem can be attacked in two ways - the so-called test can be done 
either by one experimentalist or by two experimentalists described below.:\\

\noindent
Cheating free test by one experimentalist:\\
1. Alice sends a particular
 sequence of EPR particles 
 to Bob.\\
2. Taking half of the particles,
 Bob measures the  EPR correlation.\\
3. If he gets perfect correlation then, 
with remaining particles he can himself perform
the test at two distant corners of a big laboratory.\\

This  protocol can serve   as  quantum cryptosystem if  
two shared arrangements of entangled states are used.
If two secret arrangements are used then
Bob cannot get perfectly correlated EPR data due to eavesdropping
even when Alice is honest.\\

\noindent
Cheating free test by two distant experimentalists:\\
1. Alice  sends a sequence of EPR particles from each EPR pair
to Bob.\\
2.  Bob  seeks  some of the partner EPR particles from Alice.\\
3. Alice sends particles to Bob according to his demand.\\
4. Bob measures the EPR correlation on those pairs.\\

 If he gets perfect correlation in EPR data 
(statistics should be high), he can trust the other particles.
Alice cannot deceive Bob as he does not know which particles (events) 
will 
be sought and which basis will be chosen to measure the 
EPR correlation by Bob. If Alice sends "fake" states 
in the first round
she could not create entangled state in second round.
Therefore,  genuine EPR states can be shared in this way.
(They can  use them for secure ideal quantum coin tossing which is also believed
to be impossible [20]). 
They can proceed for the inequality test with 
these remaining shared EPR pairs. But there is a loophole. They can cheat each other when they reveal results
and basis/angles of measurements.
After Alice's discloser of results and angles of measurements,
Bob can easily reveal "fake" results and the corresponding "fake" 
angles of measurements
with out going through any measurements in order 
to show violation or no violation
of Bell's inequality to Alice.
Same thing Alice can do if Bob reveals results first.
Cheating -free test can be executed by additional steps.\\

After being  sure they share genuine entangled states, the protocol can be
extended in the following manner:\\
5. Alice will ask Bob to reveal the results and
basis of measurements of half of his genuine particles.\\ 6.
 Bob meets the demand of Alice.\\
7. Getting the information from Bob, 
Alice  chooses some particles to
test the EPR correlation.
8. If she gets perfect correlation, she proceeds for the inequality test.\\

Note that  Bob does not know which event and basis will be chosen by Alice to
measure the EPR correlation. That's why the test will be a genuine test.
Similarly  Bob can be sure about the validity of the test just by asking
Alice to reveal the results and basis of measurements
of the remaining shared particles. Of course this time Bob will not
reveal anything but only measure according to the revealed data.
It is trivial to mention that the cheating-free test is simply
 a cheating-free test not 
 experimentally  loopholes-free test which is yet to performed.
In the light of cheating-free test, the issue of 
loopholes- free test can be investigated.\\

\noindent
Note added: So far we didn't consider noise.
Security of our alternative  QKD protocols 
in presence of noise is an open and interesting problem.\\

\noindent
{\bf Acknowledgement:} I wish to acknowledge the computer facility of
Variable Energy Cyclotron Center and library facility of Saha Institute
of Nuclear Physics.\\

\end{document}